\RequirePackage{lineno}
\documentclass[11pt,jkps,preprint,fleqn,showpacs,showkeys]{revtex4}
\usepackage[pdftex]{graphicx}
\usepackage{amssymb}
\usepackage{amsmath}
\usepackage{bm}
\usepackage[fleqn]{nccmath}
\usepackage{caption}
\usepackage{tabularx}
\usepackage{esvect}
\MakeRobust{\vv}
\usepackage{subfigure}
\usepackage{multirow}
\newcolumntype{b}{ >{\centering\arraybackslash}m{1in}}
\newcolumntype{s}{ >{\centering\arraybackslash}X }
\begin{document}
\setcounter{page}{0}
\title[]{A development of a compact software package for systematic error studies in muon $g-2$/EDM experiment at J-PARC}
\author{E. \surname{WON}}
\author{WOODO \surname{LEE}}
\email{woodolee@korea.ac.kr}
\thanks{Tel: +82-10-7517-2711}
\affiliation{Department of Physics, Korea University, Seoul
02841, Korea}


\begin{abstract}
The J-PARC muon g-2/EDM experiment aims to measure the muon magnetic moment anomaly $a_{\mu} = (g -2 ) / 2 ~ $.
The target sensitivity for $a_{\mu}$ is a statistical uncertainty of $450 \times 10^{-9}$, and $a_{\mu}$ can be extracted by measuring the spin precession of muons ($\omega_a$).
The $\omega_a$ can be extracted from the distribution of muon decay time, and the time is measured by reconstructing tracks of positrons which are decayed from the muons.
To extract $\omega_a$ precisely, the systematic effects on $\omega_a$ should be controlled under the aforementioned sensitivity.
To study systematic effects, we develop a compact simulation package, and using our developed package, we study several systematic effects and discuss results.
We discuss the package details and studies of systematic effects on $\omega_a$ based on our package. 
\end{abstract}



\maketitle

\section{INTRODUCTION}

The J-PARC muon $g-2$/EDM experiment \cite{10.1093/ptep/ptz030} aims to measure the muon magnetic moment anomaly $a_{\mu} = (g -2 ) / 2 ~ $ and the muon electric dipole moment (EDM) $d_{\mu}$ with a high sensitivity.
The target sensitivity is 1.0 parts per million (ppm) as an initial goal, and the ultimate goal is 0.1 ppm.
The $a_{\mu}$ is related to $\omega_a$, which can be extracted by the distribution of muon decay time.
A muon decay generates one positron and two neutrinos, and the decay time of muon can be estimated by reconstructing positron track.
The sensitivity is related to a number of reconstructed positrons, and a myriad of the positrons are required to study the systematic error effects on $\omega_a$ under the aforementioned sensitivity.
A software package is developed to simulate the systematic effects, and
such effects are studied under 10 ppm sensitivity with $10^{9}$ positrons.
The four systematic error effects on $\omega_a$ are studied with the package, respectively.
We discuss the package in detail and the systematic error effects on $\omega_a$ using our package in the following sections.

\section{The estimation of Statistical Error}

The muon magnetic moment anomaly $a_{\mu} = (g -2 ) / 2 $ can be described with five parameters.

\begin{ceqn}
\begin{align}
    a_{\mu} &= \frac{R}{\lambda - R}, \\
    R &= \frac{\omega_a}{\omega_p}, \\
    \lambda &= \frac{\mu_{\mu}}{\mu_{p}},
\end{align}
\end{ceqn}
where 
$\omega_p$ is free-proton precession frequency. 
The $\lambda$ is related to magnetic moment of muon $\mu_{\mu}$ and that of proton $\mu_{p}$.
The $\lambda$ to be measured as 3.188334513(39) or 120 ppb statistical error from muonium hyperfine structure measurement \cite{PhysRevLett.82.711}. 
The ratio $R$ was measured to be 0.0037072064(20) at BNL E821 experiment \cite{PhysRevD.73.072003}. 
The uncertainty of muon anomalous magnetic moment $a_{\mu}$ is described as 
\begin{ceqn}
\begin{align}
    \frac{\Delta a_{\mu}}{a_{\mu}} = \frac{\lambda}{\lambda - R}\sqrt{\Big(\frac{\Delta \omega_a}{\omega_a}\Big)^2 + \Big(\frac{\Delta \omega_p}{\omega_p}\Big)^2 + \Big(\frac{\Delta \lambda}{\lambda}\Big)^2}.
\end{align}
\end{ceqn}
Each systematic uncertainty on $\omega_a$, $\omega_p$, and $\lambda$ is directly propagated to the uncertainty on $a_{\mu}$, and we discuss the systematic uncertainty on $\omega_a$ later. 
When muons are under magnetic field $\overrightarrow{B}$ and electric field $\overrightarrow{E}$, the spin precession vector of muons of $\overrightarrow{\omega_a}$ can be described as \cite{spin}
\begin{ceqn}
\begin{equation}\label{eq:omega}
    \overrightarrow{\omega_a}= - \frac{e}{m_{\mu}} \left[ a_{\mu} \overrightarrow{B} - \left(a_{\mu} - \frac{1}{\gamma^2 - 1   }\right)\frac{\overrightarrow{\beta}\times \overrightarrow{E}}{e} \right], 
\end{equation}
\end{ceqn}
where $e$ is electric charge, 
$\gamma$ and $\overrightarrow{\beta}$ are the Lorentz factor of muons, 
and $m_{\mu}$ is the muon mass.

The J-PARC muon $g-2$/EDM experiment is designed to have no electric field, and therefore Eq. (\ref{eq:omega}) can be simplified to be
\begin{ceqn}
\begin{equation}\label{eq:omegaB}
    \overrightarrow{\omega_a} = \frac{e}{m_{\mu}}a_{\mu}\overrightarrow{B}.
\end{equation}
\end{ceqn}
We cannot measure $\omega_a$ directly, but it can be obtained from the distribution of the muon decay time.
From the number of muon decays in a certain time interval, we can calculate the required number of positron to attain a statistical error of $\omega_a$.
In the ideal case, the number of muons at a time ($t$)  can be described as
\begin{ceqn}
\begin{equation} 
\centering
    N_0 e^{-t/\gamma \tau} 
    [
        1- A\cos{(\omega_a t + \phi)}
    ],
    \label{eq:decaypdf}
\end{equation}
\end{ceqn}
where $N_0$ is the number of initial muons,
$\tau$ is the muon mean lifetime, 
$A$ is the normalization to the oscillation term
of the decay positron, and $\phi$ is the initial phase. 
Here $N_0$ and $A$ can be changed depending on the threshold energy of the positron, and the extraction of the $\omega_a$ is a 5-parameter fit to the positron time distribution.
The expected statistical error normalized on $\omega_a$ ($e_{\omega_a}$) can be expressed by \cite{bnl},
\begin{ceqn}
\begin{eqnarray}
e_{\omega_a} = \frac{\sqrt{2}}{{\omega_a}\gamma \tau A \sqrt{N} }, 
\label{eq:err}
\end{eqnarray}
\end{ceqn}
where $N$ is the total number of entries (total number of positrons) in the binned histogram to fit.
Note that the sensitivity is directly related to the $N$.

\section{Software package}
We develop a new software package to study the systematic effects of the experiment.
There are three stages in the package, as we discuss below.
First, Positron Generator (PG) is developed to reduce the positron generation time significantly by using the analytic formula of the muon decay.
The PG generates the information of positron when muons are decayed according to the weak decay of three particles.
Second, the generated positrons are simulated by GEANT4 \cite{Geant4} to obtain hit information on the detector.
Third, positron tracks are reconstructed using the GENFIT toolkit \cite{genfit2}, which is based on Kalman Filter (KF) \cite{kf}, and the decay times of muons are also reconstructed in this step.

\subsection{Step 1. Positron generation}

The GEANT4 program can, in principle, be used in the simulation of muon decay and spin precession in order to obtain the positron information from a muon decay.
However, the numerical calculation for every step with the spin tracking causes a very long calculation time.
The GEANT4 simulation takes 61 seconds to make 4 $\times ~ 10^4$ positrons with 1 single CPU of 3.7 GHz speed.
Note that one bunch contains 4 $\times ~ 10^4$ muons that come in every 40 ms \cite{10.1093/ptep/ptz030}.
To reduce the simulation time, our PG is developed.
The PG takes only 10 seconds with the same CPU to generate 4 $\times ~ 10^4$ positrons by using the analytic formula according to the weak interaction to generate positrons.
The true muon decay time ($t_{true}$) and the position of the muons at that moment are calculated by the PG.
Since the magnitude of $\overrightarrow{B}$ is 3 T and the muon momentum is 300 MeV/$c$, the muon will be on a circle of the radius 333 mm, and therefore the position of decay vertex of the positron will be on this circle.
In order to compute the momentum of the positron ($p_{e^+}$), the PG uses an accept-rejection Monte Carlo method, which follows the Michel Spectrum \cite{michel}, and the spin direction is also calculated.
The positrons are generated with the magnitude of the muon momentum of 300 MeV/$c$.

\begin{figure}[h]
    \centering
    \includegraphics[width=\textwidth]{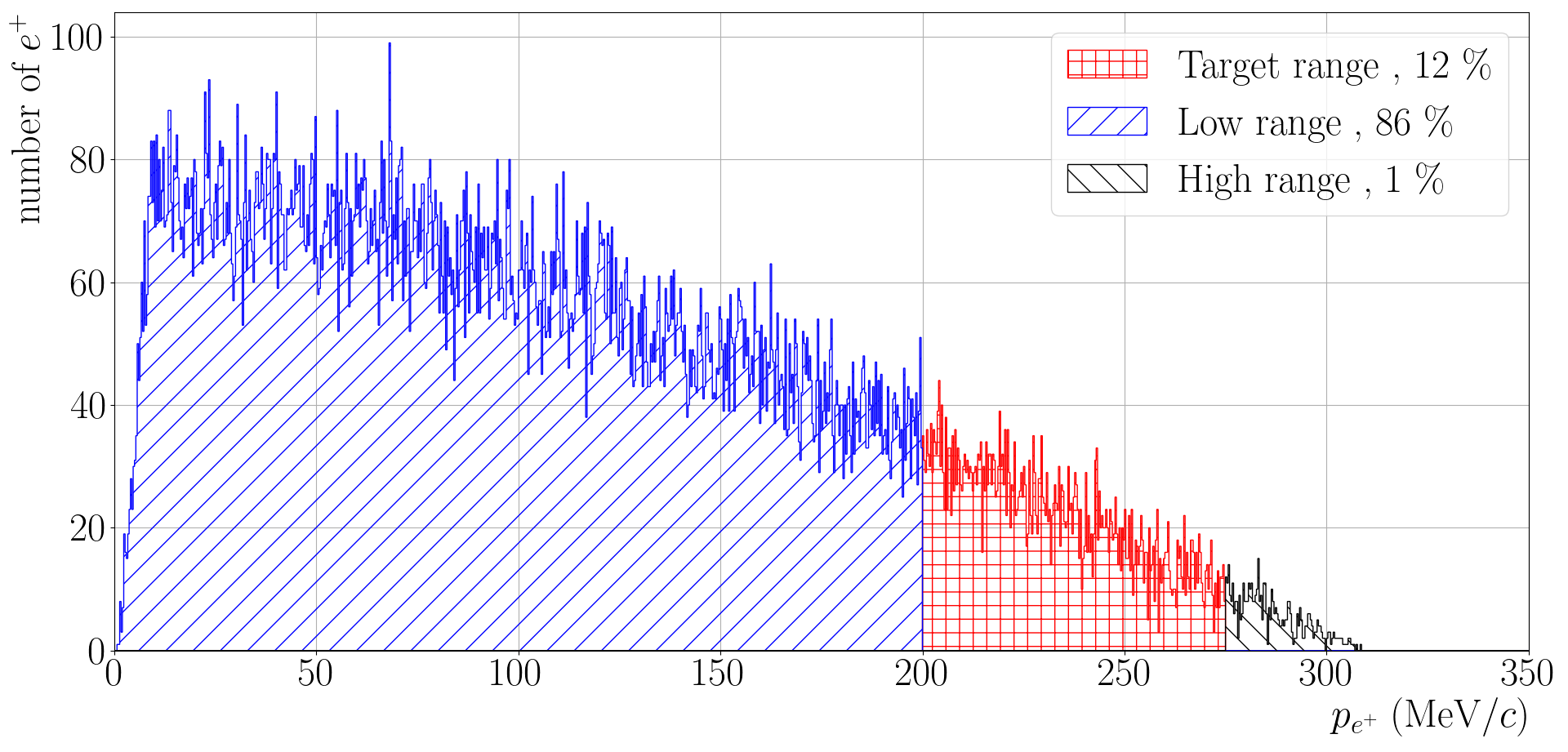}
    \caption{Positron momentum distribution from PG is shown. 
    }
    \label{fig:momentumRange}
\end{figure}

Figure \ref{fig:momentumRange} shows the distribution of generated $p_{e^+}$. 
In total, $4 \times 10^4$ of $p_{e^+}$ are shown in here.
We require the momentum of the positron to be 200 to 275 MeV/$c$ in the generation of the positron since this selection will be required at the offline stage.
This selection is based on a full GEANT4 based generation, finding, and fitting positron track studies \cite{tdrFull}.
After the momentum selection, about 12 \% of events are accepted as signal events.
Our PG is used in the next stage of the detector simulation.

\subsection{Step 2. Detector simulation} \label{detector}
The GEANT4 package is used for simulating the
interaction of the positron with the detector to obtain the information of positron hits.
The generated positrons from the PG are simulated with the detectors by the GEANT4 simulation.
In total, 40 detectors are placed in the simulation, and the detector geometry is prepared with Geometry Description Markup Language (GDML) \cite{gdml} format.
\begin{figure}[h]
    \centering
    \includegraphics[width=\textwidth]{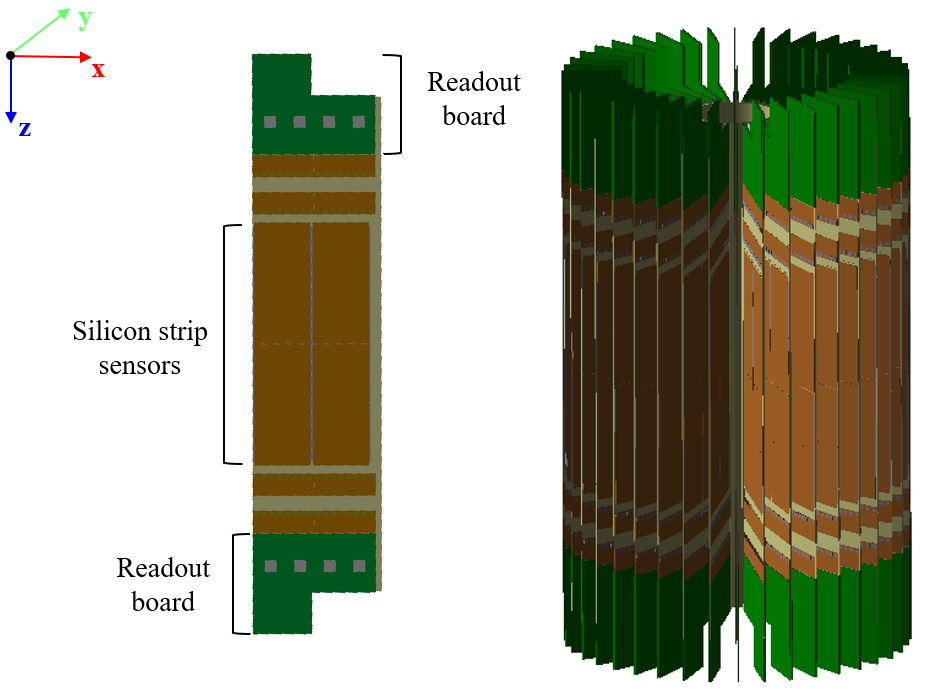}
    \caption{The positron detector is shown. 
    The positron detector consists of in total 40 modules.
    One module contains four arrays of silicon strip sensors. 
    }
    \label{fig:vane}
\end{figure}
In total, 40 detector modules are mounted in radial direction like vanes in the experiment \cite{detector}.
Figure \ref{fig:vane} shows the geometrical setup the modules.
The central part consists of the strip silicon sensor, and two readout systems are placed at the top and bottom position of silicon strip sensor arrays, and the detector modules are placed on a vacuum and a 3 T static magnetic field.

\begin{figure}[h]
    \centering
    \includegraphics[width=\textwidth]{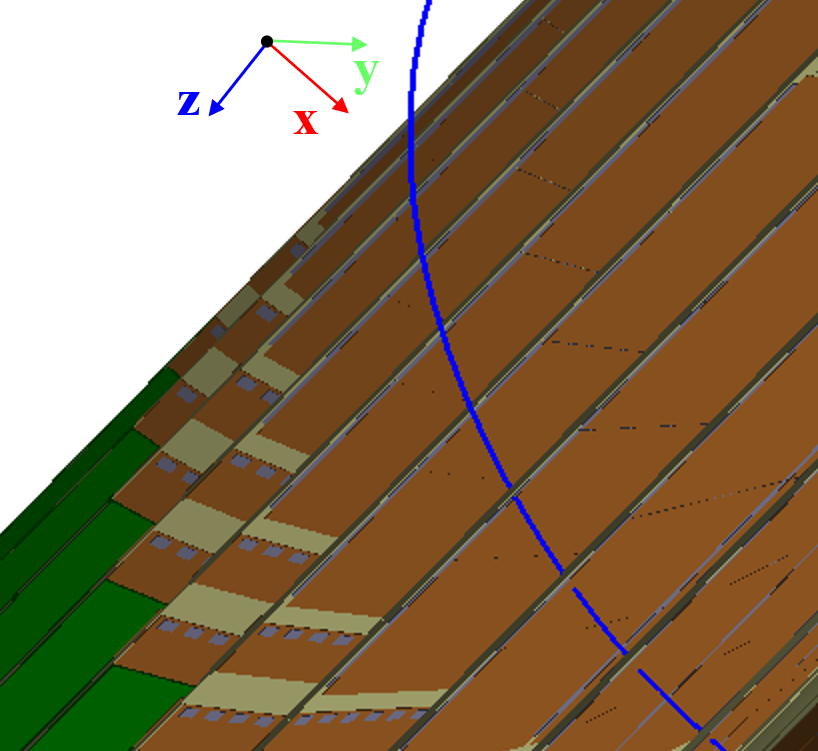}
    \caption{A zoomed view of one event is shown.
        The distance from the vane center to the muon orbit is $333$ mm, and a muon is decayed in the orbit.
        For this event, the generated muon has the decay time of 3.61 $\mu$s, and 
        with the positron with the momentum 253.53 MeV/$c$.
        }
    \label{fig:g4}
\end{figure}

Figure \ref{fig:g4} shows an example of our simulation.
The positron in the magnetic field follows a helix trajectory, hits the detectors, and finally annihilates.
All the information of positrons, including time, positions, and momenta, are obtained from the simulation. 
Instead of implementing silicon strip sensors in the GEANT4, we implement plane sensors without strips, which reduces the storage for saving the information from a myriad of strip sensors.
Considering the gap between adjacent two strip sensors, which is 190 $\mu$m, the hit positions are smeared with the Gaussian distribution.
The $\sigma$ of the distribution is set to be $190 \slash \sqrt{12} = 54 ~ \mu$m, and the hit points are reconstructed by calculating the middle points of the information from both sides of the detector.

To minimize the simulation time, we stop the GEANT4 simulation after obtaining enough information to reconstruct the tracks.
In principle, five reconstructed hit points are required to fit the track with the helix parameters.
This makes the minimum number of the points of the track.
However, we require two more points because some events hit the silicon detectors only one side of them, and in that case, the hit point cannot be reconstructed.
When the total number of hits from both radial and vertical sensors becomes fourteen, we stop the simulation.

\subsection{Step 3. Track fitting} 
In this section, we discuss the fit program of our package.
We reconstruct the momentum of positrons from the simulated hit information to estimate the decay time of muons in this step. 
As the trajectory of a positron in a magnetic field is a helix, the fitter requires a minimum of five parameters \cite{helix}.
We use GENFIT, which is based on KF and Deterministic Annealing Filter (DAF) \cite{daf}, for fitting the momentum of positrons when muons are decayed.
For fitting the track, the initial states of each track should be set up, and the initial states consist of the momentum and the position of the positron when the positron hits the detector first.
The initial momentum of the positron is obtained by fitting the circle, which is drawn with the first three hits, and the initial position is from the first hit of the positron. 

\begin{figure}[h!]
    \centering
    \includegraphics[width=1.1\textwidth]{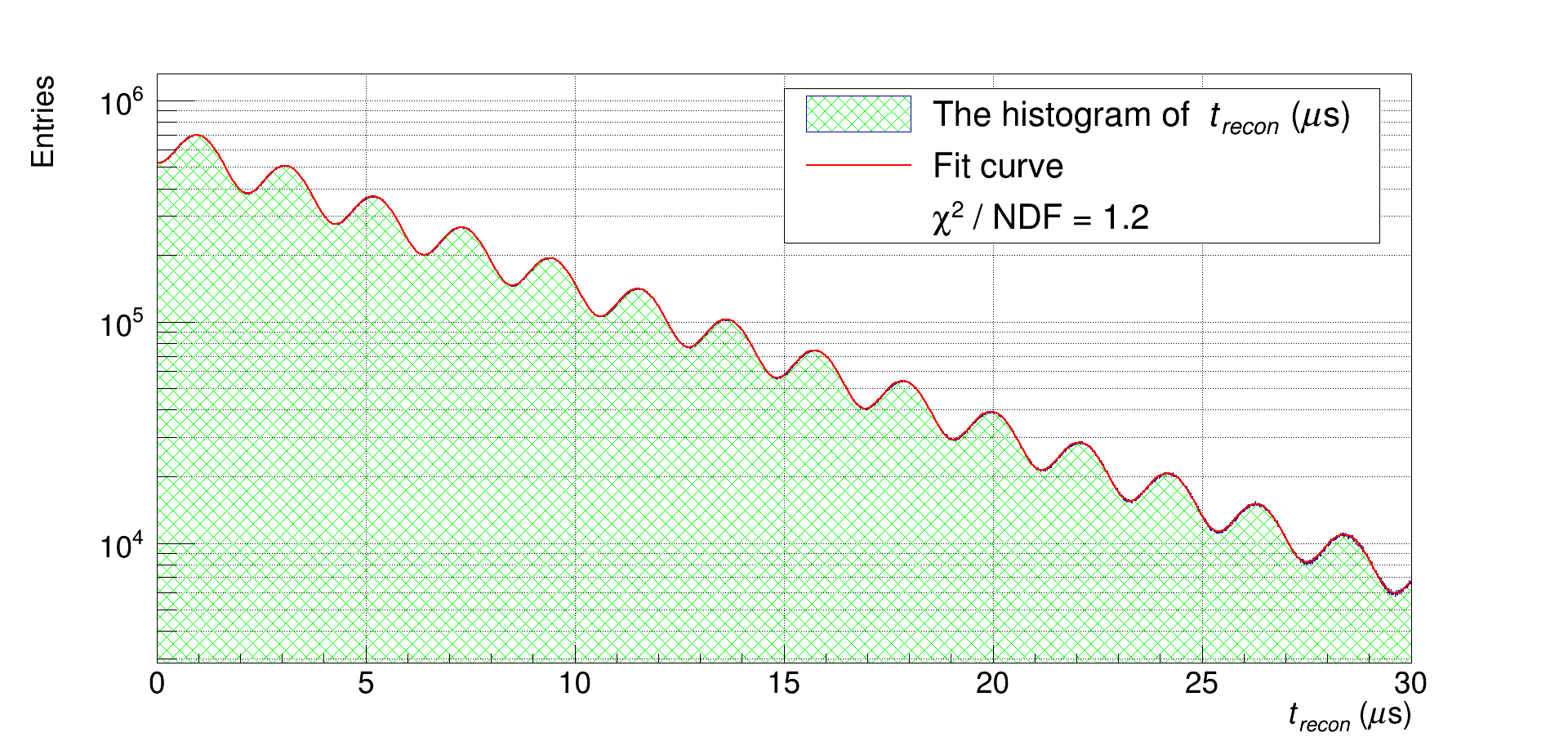}
    \caption{The distribution of $t_{recon}$ is shown with the hatched histogram (green), and the fit curve (red) is shown.
    The $\chi^{2}$ / number of degrees of freedom (NDF) is 1.2.}
    \label{fig:fittingResult}
\end{figure}

For validating the performance of the fit program, we check the momentum resolution ($R_p$), which is defined as 
\begin{ceqn}
\begin{equation} \label{resolution}
\centering
R_p  = \frac{p_{recon} - p_{true}}{p_{true}},
\end{equation}
\end{ceqn}
where $p_{true}$ is the true momentum from the PG, and $p_{recon}$ is the reconstructed momentum.
Figure \ref{fig:fittingResult} shows the distribution of $R_p$, and $90 ~ \%$ of events are within $| R_p | < 0.1 $.

\begin{figure}[h]
    \centering
    \includegraphics[width=\textwidth]{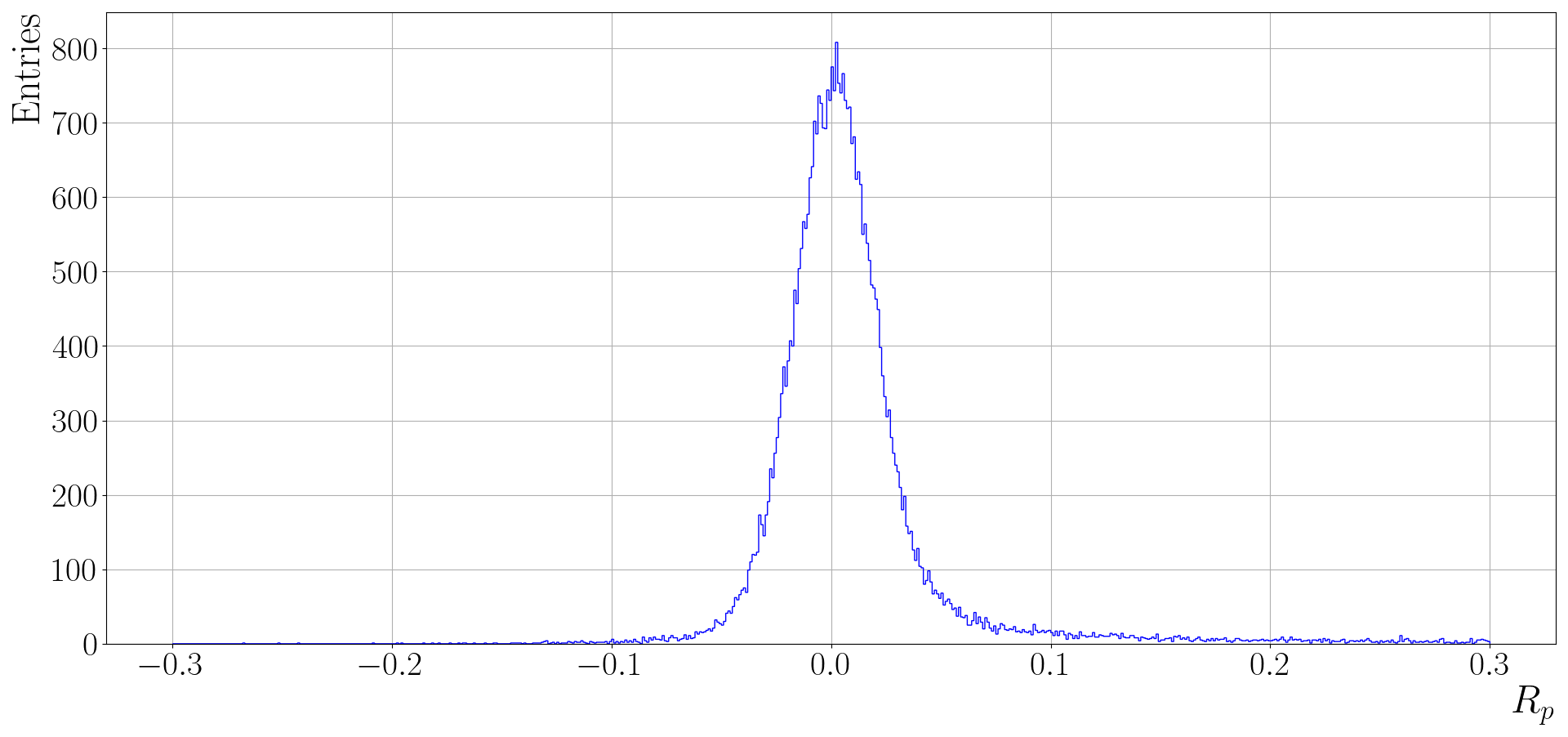}
    \caption{The distribution of $R_p$ is shown.
    In total, $4\times10^4$ of the reconstructed tracks are shown.
     }
    \label{fig:genfitResult}
\end{figure}

The muon decay time is computed by back-tracking from the first hit time and the $p_{recon}$.  
At the end, we require $p_{recon}$ to be within 200 to 275 MeV/$c$.
The reconstructed muon decay time ($t_{recon}$) is used to extract $\omega_a$, which is discussed in the next section.

\section{The extraction of $\omega_a$}
The $\omega_a$ is directly related to the anomalous magnetic momentum of the muon, and it can be extracted from the fit to the distribution of $t_{recon}$.
The fit function has five parameters, based on Eq. (\ref{eq:decaypdf}).
Here, we compare two values of $\omega_a$ where one is from the fit to the distribution of $t_{true}$, and the other from the distribution of $t_{decay}$.

Table \ref{table:fittingResult} shows the results.
The polarization factor ($P$) of the muon is assumed to be 0.5.
The relative statistical uncertainties with respect to the fit value for $\omega_a$ ($e_{\omega_a}$) are found to be 13 ppm, and it is larger than the small difference in the central values of $\omega_a$ from fits using $t_{true}$ and $t_{recon}$.
This demonstrates our simulation packages do not introduce any unwanted bias at this state.

\begin{table}[h!]
\begin{tabularx}{\textwidth} { 
   >{\centering\arraybackslash}X 
   >{\centering\arraybackslash}X 
   >{\centering\arraybackslash}X
   >{\centering\arraybackslash}X
   >{\centering\arraybackslash}X
   >{\centering\arraybackslash}X
   }
 \hline
 Cases & $PA$ & $\gamma \tau$ & $\omega_a$ & $\phi$ & $e_{\omega_a}$ (ppm)\\ 
 \hline
  $t_{true}$ & 0.214356(47) & 6.60029(25) & 2.976413(39) & $-0.00112(33)$ & 13 \\
  $t_{recon}$  & 0.214367(47) & 6.60038(25) & 2.976411(38) & $-0.00024(33)$ & 13 \\
 \hline
\end{tabularx}
\caption{The fit result of five parameters with $t_{true}$ and $t_{recon}$, respectively.
}
\label{table:fittingResult}
\end{table}

Here, we discuss the required time and storage to reach the target sensitivity using the fit procedure above.
Table \ref{table:resources} shows the sensitivity, 
the number of generated positrons, the running time ($t_{r}$) of the simulation, and the required size of storage.
The simulation is carried out with 253 threads on a 2.5 GHz clock speed CPU.
To achieve 1 ppm sensitivity, it is expected that at least 404 days and 242 TB are required.
Since the study of the systematic effects on $\omega_a$ down to 1 ppm is unrealistic with the computing resources that we have, we study them with $10^{9}$ positrons.

\begin{table}[h!]
\begin{tabularx}{\textwidth} { 
   >{\centering\arraybackslash}X 
   >{\centering\arraybackslash}X 
   >{\centering\arraybackslash}X 
   >{\centering\arraybackslash}X }
\hline
$e_{\omega_a}$ & $N$ & $t_{r}$ & Storage \\ 
 \hline
 13 ppm  &    1.2 $\cdot10^{9}$ & 48 (hours)  & 1200 (GB)\\
 1 ppm  &    1.2 $\cdot10^{12}$ & 404 (days)  & 242 (TB)\\
 \hline
\end{tabularx}
\caption{
The expected values of $e_{\omega_a}$, $N$, $t_r$, and the size of data storage are summarized.
}
\label{table:resources}

\end{table}

\section{Systematic errors}
Our simulation package is utilized for the study of systematic error effects on $\omega_a$.
We discuss the sources of potential systematic error effects on $\omega_a$ in the following sections, and four sources we consider in this paper are shown in Table. \ref{table:sysError}.

\begin{table}[h!]
\begin{tabularx}{\textwidth} { 
   >{\centering\arraybackslash}X |
   >{\centering\arraybackslash}X 
   }
    \hline
 Sources  & Description  \\ 
 \hline
 Timing shift due to pile-up & Timing shift effects due to the piling up events.\\
  \hline
 Electric and magnetic field &   The effect from the uncertainty of  Electric and magnetic field.\\
  \hline
 High energy positron &   The effect on high energy positrons.\\
  \hline
 Differential decay & The muon momentum spread effect. \\
 \hline
\end{tabularx}
\caption{
Four sources of systematic effects that we study and their descriptions are shown.
}

\label{table:sysError}
\end{table}

First, we study the effect of piled-up positrons on the $\omega_{a}$.
When a bunch of the muon decays in the storage area, some positrons arrive at the detectors at the same time and in the same place, which is piled up. 
The pile-up time can be limited to be distinguished from each other due to the limitation of the readout system \cite{detector}, and $\omega_{a}$ is shifted by this effect.
Second, the shifted $\omega_{a}$ due to the uncertainty of the electromagnetic field is studied. 
In principle, the electric field is zero, and the magnetic field is constant as 3 T in the muon storage.
However, a non-zero electric field and the unstable magnetic field exist, and $\omega_a$ is shifted due to them. 
Third, the effect from high energy positron to $\omega_{a}$ is studied. 
Lastly, we study the effect on muon momentum spread in the differential decay section, which comes from the muon beam's uncertainty itself.

\subsection{Timing shift effect due to pile-up positrons}
The muon beam is injected into the muon storage at every 40 ms at J-PARC muon beamline.
One bunch has in total 4 $\times ~ 10^{4}$ muons ($N_b$). 
When multiple positrons arrive at the detector within a short time period, hit times and tracks of such positrons may not be correctly reconstructed.
In this case, detected hit times from multiple positrons may be in the same time bin of the $\omega_a$ histogram where one bin corresponds nominally 5 ns.
This may lead to a bias in the extraction of $\omega_a$.
In principle, the effect of this can be done by injecting multiple positrons in the GEANT4 setup, pass the hit information to the finder and fitter, and do the $\omega_a$ fit.
However, at this moment, there is no finder available and because of that, we use the following simplified method.

\begin{figure}[h]
    \centering
    \includegraphics[width=\textwidth]{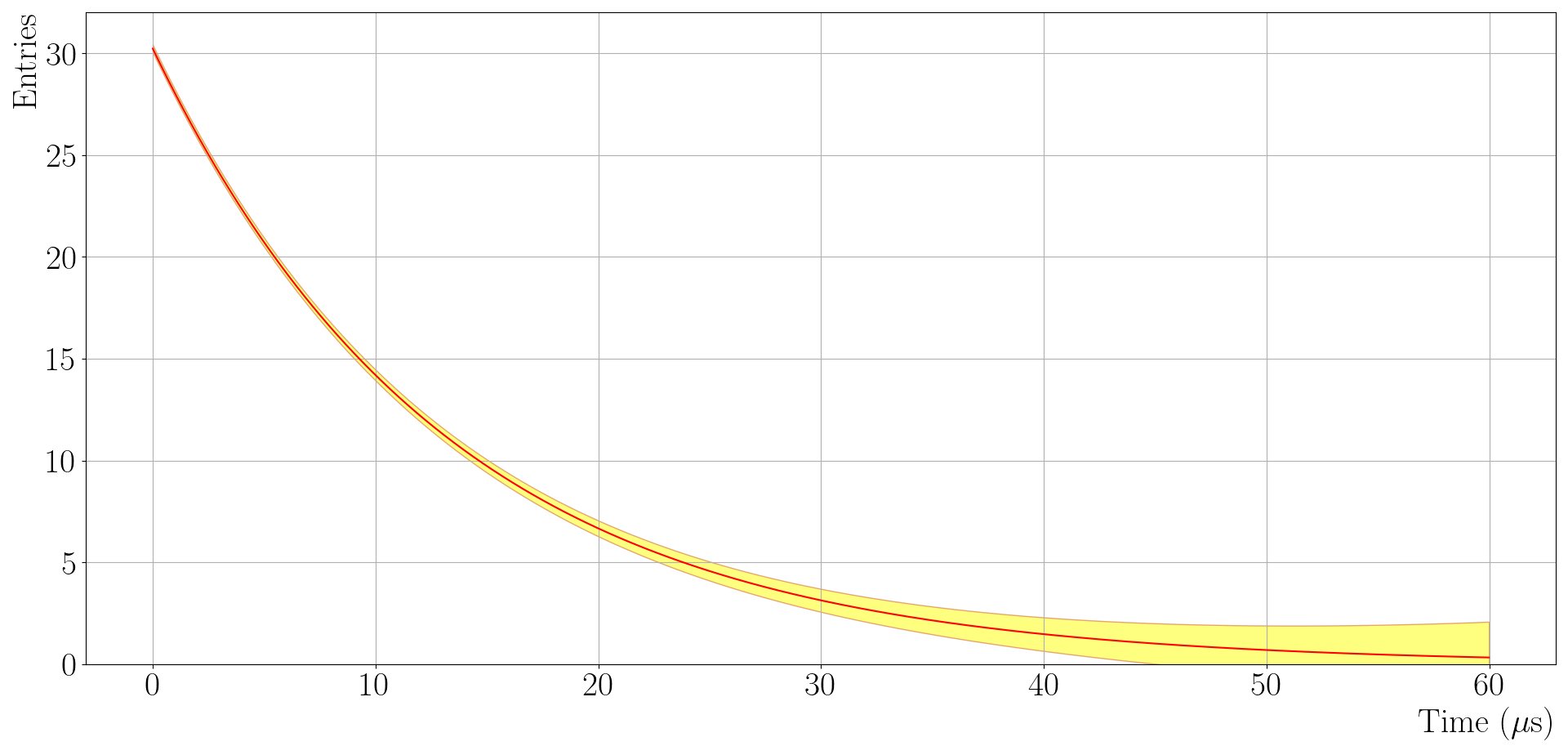}
    \caption{The number of positrons for each 5 ns time window with $N_b$ positrons as a function of the decay time is shown.
    The solid line (red) shows the number of detected positrons in each time bin, and the area (yellow) indicate size of statistical uncertainty.
    In total, 30 positrons ($n_b$).
    }
    \label{fig:pileuprate}
\end{figure}

We generate $N_b$ positrons and make a histogram with 5 ns bin because the timing resolution of the detector corresponds 5 ns \cite{detector}.
Figure \ref{fig:pileuprate} shows the histogram with $N_b$ positrons.
We assume the maximum number of multiple positrons in a 5 ns window in a bunch is the number of entries in the first bin in such histogram ($n_b$).
Then, we scale up this histogram by $N_b / n_b$.
As the last step, we shift 5 ns early or late and add it to the default histogram with a total entry of $N$.
This may underestimate the effect of pile-up and will be studied once the realistic setup is available.

Table \ref{table:pileUpresult} shows for six different cases with pile-up with $N_b$, $N_b/10$, and $N_b/100$.
The difference between the $\omega_a$ for nominal setup and that of the pile-up case normalized by nominal value $\omega_a$ are shown.
Since the statistical uncertainty is 13 ppm, it seems there is no systematic effect due to the pile-up at this stage.

\begin{table}
\begin{tabularx}{\textwidth} { 
   >{\centering\arraybackslash}X |
   >{\centering\arraybackslash}X |
   >{\centering\arraybackslash}X |
   >{\centering\arraybackslash}X |
   >{\centering\arraybackslash}X |
   >{\centering\arraybackslash}X 
   }
  \hline
      \multicolumn{2}{c|}{ $N_b$ pile-up case} &
      \multicolumn{2}{c|}{ $N_b / 10 $ pile-up case} &
      \multicolumn{2}{c}{ $N_b / 100$ pile-up case} \\
          \hline
    Early & Late & Early & Late & Early & Late \\
    \hline
     2.0 ppm & 3.6 ppm & 1.0 ppm & 1.0 ppm & 0.0 ppm & 0.3 ppm \\
    \hline
\end{tabularx}
\caption{It shows for six different cases with pile-up with $N_b$, $N_b/10$, and $N_b/100$.
The difference between the $\omega_a$ for nominal setup and that of the pile-up case normalized by nominal value $\omega_a$ are shown.}
\label{table:pileUpresult}
\end{table}

\subsection{Electric and magnetic field effects} 

The non-uniformity of the electric and magnetic field in the muon storage directly affects the orbit of muons, as well as on $t_{recon}$.
In principle, the magnetic field ($\overrightarrow{B}$) should be constant at 3 T in the experiment.
However, the deviation from the 3 T DC magnetic field
and the stray magnetic field from the electronics can exist in the experiment.

The magnitude of the inhomogeneity from the 3 T DC field is estimated to be 0.2 ppm peak-peak value according to Ref. \cite{magnet}.
This inhomogeneity can change the orbit of muons in the muon storage.
The $t_{recon}$ can, in principle, also be distorted as the result of the distorted muon orbit.
In addition, the $t_{recon}$ can be distorted when they are reconstructed because we assume that the $|\overrightarrow{B}|$ is constant (3 T) along the $z$ direction for reconstructing $t_{recon}$ from the first hit time.
We simulate the effect by adding 0.6 $\mu$T ($\delta B$) along the $x$, $y$, and $z$ coordinates of $\overrightarrow{B}$ ($B_x$, $B_y$, $B_z$), respectively.
The $e_{\omega_a}$s are obtained, and they are shown in Table \ref{table:dB}.
Note that the results from applying inhomogeneity along the $z$ direction are larger than statistical uncertainty (13 ppm).
Therefore, further efforts to reduce the inhomogeneity along the $z$ direction are anticipated in order to have a systematic effect due to the inhomogeneous magnetic field smaller than statistical uncertainty.

\begin{table}
\centering
\begin{tabularx}{\textwidth} { b |s |s |s |s |s |s }
    \hline
      \multicolumn{1}{c|}{   } &
      \multicolumn{2}{c|}{ $B_x$} &
      \multicolumn{2}{c|}{ $B_y$} &
      \multicolumn{2}{c}{  $B_z$} \\
          \hline
  The sign of $\delta B$ & $+$ & $-$ &  $+$ & $-$ &$+$ &  $-$ \\
    \hline
     $e_{\omega_a}$ & 9 ppm & 7 ppm & 10 ppm & 3 ppm & 17 ppm & 11 ppm \\
    \hline
\end{tabularx}
  \caption{The results of the simulation for the systematic error effect due to the magnetic field uncertainty are shown.}
\label{table:dB}
\end{table}

Concerning the stray magnetic field from the electronics, we consider the stray magnetic field from the inductor in the DC to DC converter.
Note that the AC of 6 A with the frequency of 970 kHz is to be applied in the inductor of the DC to DC converter.
We calculate magnetic field radiation from the dipole approximation and find that the magnitude of the magnetic field from the radiation is at the 0.03 ppb level with respect to 3 T field Ref. \cite{dc2dc}.
Therefore, we conclude that this is negligible.

The non-zero electric field can, in principle, influence the muon orbit and subsequently $\omega_a$.
Here, we examine a possible source of a non-zero electric field, again from the dipole radiation.
The expected radiation to the muon storage orbit is 0.2 $\mu$V/m and is much smaller than the requirement of 10 V/m \cite{tdrFull}.
Therefore, this source can be neglected.

\subsection{High energy positron effect}
\begin{figure*}[!t]
\centering
\subfigure[]{
\includegraphics[width=.45\columnwidth]{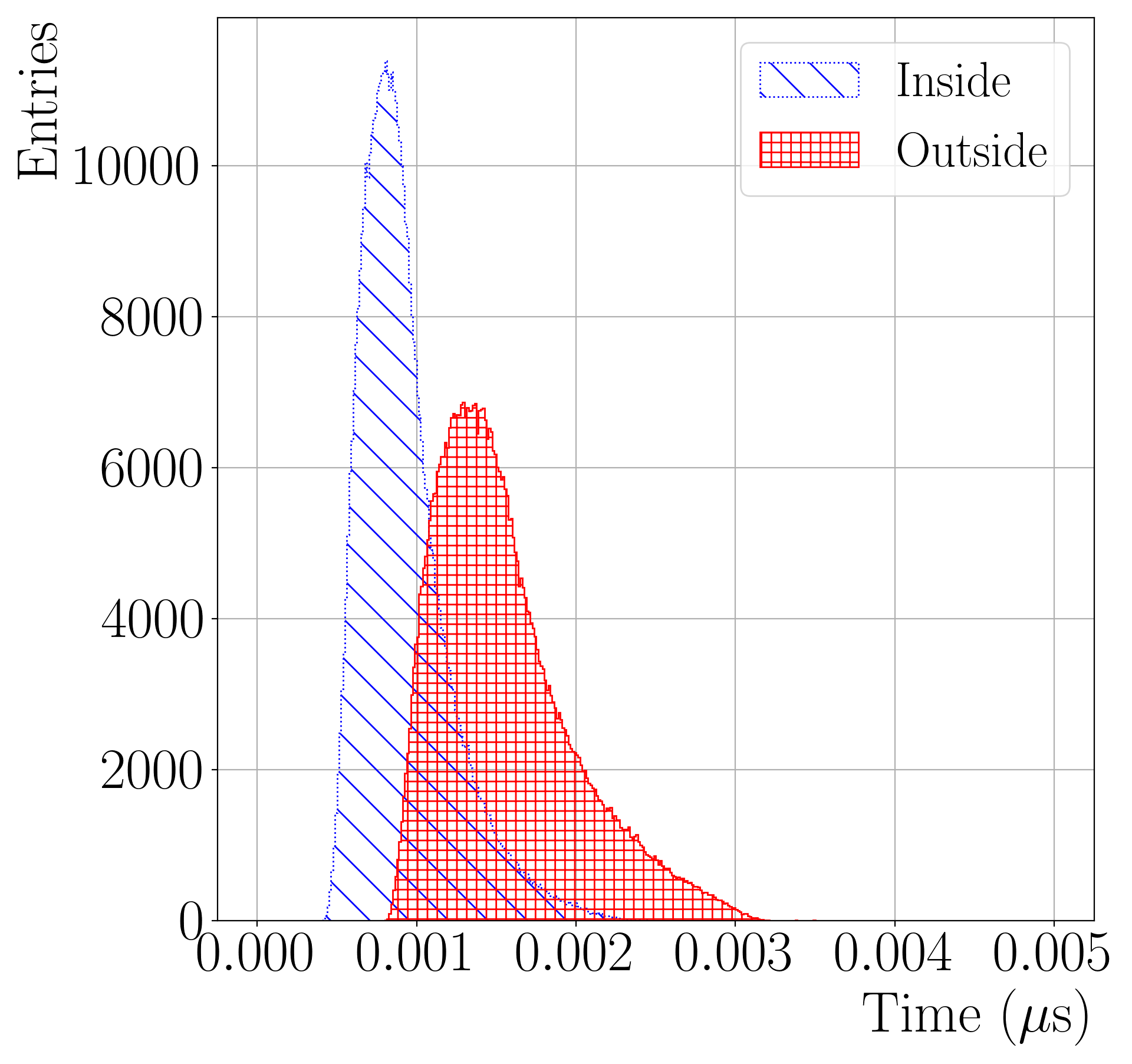}
}
\subfigure[]{
\includegraphics[width=.45\columnwidth]{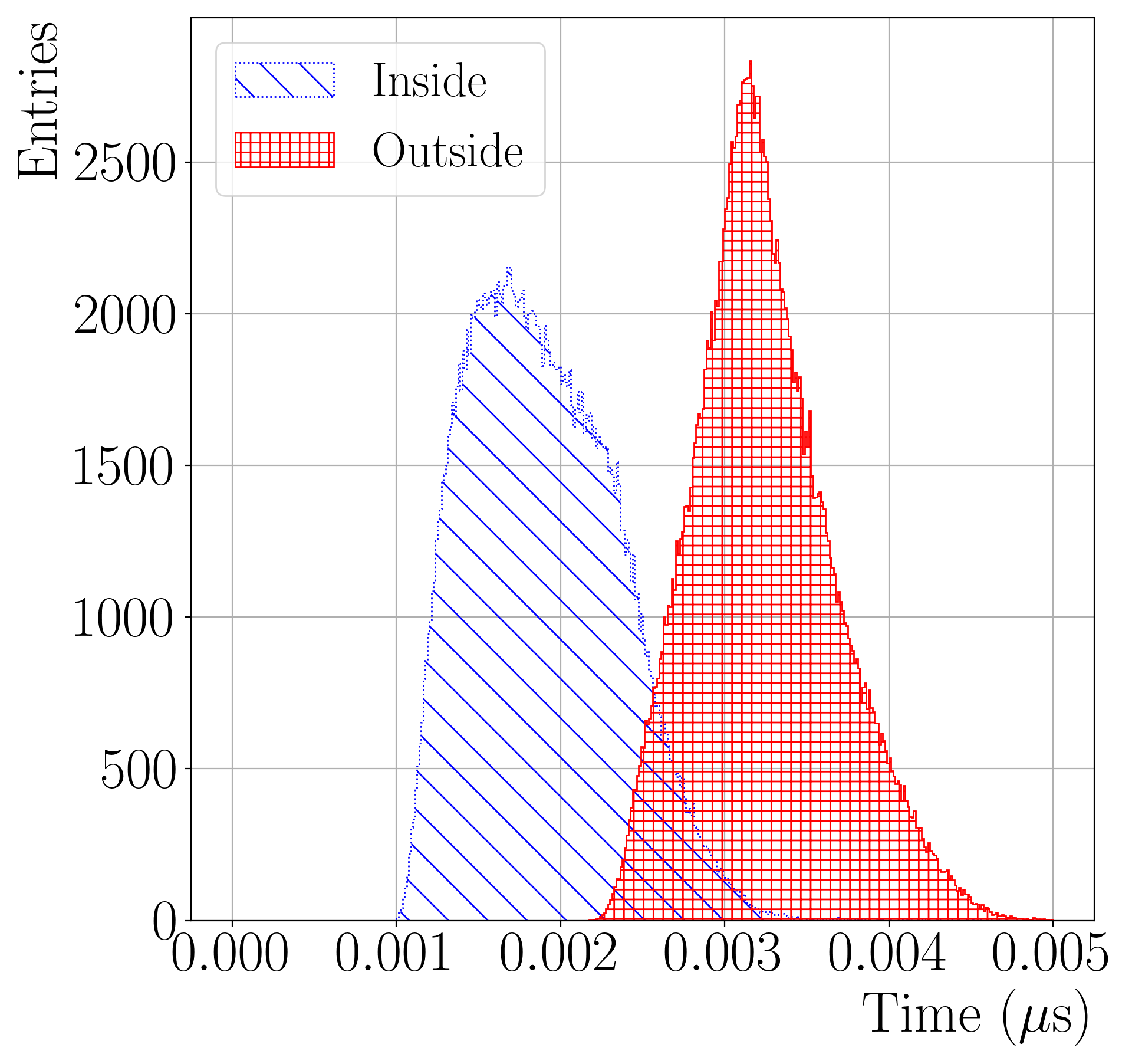}
}
\caption{
The left histograms (a) show such time difference for positrons decay direction to be inside of the detector volume (diagonaly hatched, blue) and to be outside detector volume (vertical and horizontaly hatch, red) for positrons of momentum from 200 to 275 MeV/$c$.
The right histograms (b) show similar ones for positrons of momentum greater than 275 MeV/$c$.
}
\label{fig:FlyT}
\end{figure*}

The positrons with momentum greater than 275 MeV/$c$ may not enter into the detector volume immediately depending on the direction of the decay.
In that case, such positrons travel outside of the detector volumes and hit the detector volume later.
This may cause unwanted bias in the measurement of $\omega_a$.
To study this, we look at the time between the positron decay and the first hit time for positrons of momentum from 200 to 275 MeV/$c$.
Figure \ref{fig:FlyT} (a) shows such time difference for positrons decay direction to be inside of the detector volume (diagonally hatched, blue) and to be outside detector volume (vertical and horizontally hatch, red) histogram.
Notice that there is a small shift between the two histograms.
We repeat this study on positrons with generated momentum greater than 275 MeV/$c$.
The difference becomes larger in this case, and this indicates there may be a systematic shift in the $\omega_a$ measurement due to this.
Here, to see the effect clearly, we collect the same number of positrons from two different momentum ranges.
We repeat the simulation by including positrons in the high momentum range and found that the normalized difference of $\omega_a$ between nominal and this simulation to be 1.0 ppm. 
Therefore, this effect is much smaller than 13 ppm.

\subsection{Differential decay effect}

As the final source of the systematic effect, we discuss the effect due to non-zero dispersion of the muon momentum.
Such a dispersion is known to be in the order of 0.04 $\%$ \cite{10.1093/ptep/ptz030}.  
This dispersion is directly related to the orbit of the muon, and the energy spectrum of positrons.
First, the radial distance at the muon decay can be dispersed.
Second, the momentum of positrons is also be affected.
To study these effects, we generate positrons with our PG and subsequent simulation package.
We repeat the $\omega_a$ fit with the new sample and find that the normalized difference of $\omega_a$ between nominal and this simulation to be 9 ppm.
Therefore, this effect is much smaller than 13 ppm.

\section{Conclusions} \label{results}

We develop a compact software package for fast generation, detector simulation, and track fitting of muons.
The PG part of our package is found to be six times faster than the GEANT4 based one.
The package is based on Monte Carlo simulation, GEANT4, and GENFIT for the reconstruction of positron's momentum. 
We also expect faster simulation of our package in the detector simulation stage since we parameterize the hit resolution.
From this, generation of $10^{9}$ positron decays can be carried out in less than 48 hours with 13 ppm sensitivity on $\omega_a$.
We use our compact package to study the following four systematic error effects on $\omega_a$.

\begin{table}[h!]
\begin{tabularx}{\textwidth} { 
   >{\centering\arraybackslash}X 
   >{\centering\arraybackslash}X 
   }
   \hline
 Source  & $e_{sys}$ (ppm) \\ 
 \hline
 Timing shift due to pile up   & $\leq$ 3    \\
 Magnetic field effect        & $\leq$ 17       \\
 High energy positron          & 1       \\
 Differential decay            & 9        \\
 \hline
\end{tabularx}
\caption{Each systematic error effects on the $\omega_{a}$ measurement are shown.
The fit error is 13 ppm for all.}
\label{table:sysErrorResult}
\end{table}

Table \ref{table:sysErrorResult} shows the summary of systematic error studies.
It indicates that the effect from the magnetic field is the largest source among systematic effects that we studied.
However, our systematic studies are carried out with certain assumptions, and therefore in the future for realistic studies should also be carried out to confirm our conclusions.

\section{Acknowledgements}
We would like to sincerely thank Institute for Nuclear and Particle Astrophysics (INPA) at Seoul National University.
This work is supported by a Korea University Grant, the National Research Foundation of Korea Grant funded by the Korean government (Grant No. NRF-2017R1A2B3007018), and BK21 FOUR program at Korea University, $Initiative\, for\, science\, frontiers\, on\, upcoming\, challenges$.
We also appreciate Tsutomu Mibe, Takashi Yamanaka, Yutaro Sato, and the members of the muon $g-2$/EDM experiment at J-PARC collaboration.

\end{document}